\documentstyle[aps,multicol,epsfig%
]{revtex}   

\newcommand{\beq}{\begin{equation}}  
\newcommand{\eeq}{\end{equation}}  
\newcommand{\bea}{\begin{eqnarray}}  
\newcommand{\eea}{\end{eqnarray}}  
  
\begin{document}     
\title{Classical analysis of correlated multiple ionization in strong fields}
\author{
Bruno Eckhardt$^1$ and Krzysztof Sacha$^{1,2}$
}     
\address{$^1$ Fachbereich Physik, Philipps Universit\"at    
	 Marburg, D-35032 Marburg, Germany}    
\address{$^2$ Instytut Fizyki im. Mariana Smoluchowskiego,
Uniwersytet Jagiello\'nski, ul. Reymonta 4,
PL-30-059 Krak\'ow, Poland}
\date{\today}     
\maketitle{ }   
  
\begin{abstract}
We discuss the final stages of the simultaneous ionization of two or
more electrons due to a strong laser pulse. An analysis of the classical
dynamics suggests that the dominant pathway for non-sequential escape
has the electrons escaping in a symmetric arrangement. Classical
trajectory models within and near to this symmetry subspace
support the theoretical considerations and give final momentum
distributions in close agreement with experiments.
\end{abstract}     
  
\begin{multicols}{2}

\section{Introduction}
Classical models for atomic processes can provide useful insights
in situations where quantum effects are not prominent, as for instance 
in the dynamics of highly excited states or in multiphoton 
absorption processes. 
Microwave ionization of hydrogen and other atoms
with a single valence electron\cite{koch1995,delande1991,dz1997}, 
localized wave packet dynamics or 
electron scattering off atoms come to mind \cite{Rost}. They provide a natural 
starting point for semiclassical investigations that include, at
least approximately, quantum interference effects. In particular
the ionization of Rydberg atoms in linear and elliptically
polarized microwave fields has received considerable attention and 
similarities and differences between classical and quantal behaviour
have been sorted out in great detail \cite{koch1995,delande1991,dz1997}. 
In most cases only a
single electron is considered and sufficient to interpret the observations.

Interactions between electrons seem to play an important role
in multiphoton multiple ionization in strong laser fields.
Experiments show that the yield of multiply charged ions is much higher
than can be expected on the basis of an independent electron 
model \cite{Kulander1,Kulander2}.
More recently, it has been noted that the electrons can escape
non-sequentially and that they are correlated in their final state
\cite{weber1,weber2,rottke,weber3}.
This correlation in the final state came as a big surprise and it is
our main objective here to discuss a classical model for it.

\section{The model}

The full process of multiphoton multiple ionization is quite complicated
and involves many steps. A plausible model relevant for the field
intensities of the experiments is the rescattering mechanism 
\cite{Corkum,Kulander3,Kulander4,becker1,becker2,Kulander5,becker3,becker4}. 
Before the
pulse arrives, the atom is in its ground state. Then one
electron escapes from the atom, most likely by tunnelling through
the Stark barrier. This electron is then accelerated by the field
and can be reflected back towards the atom. During this impact energy
is transfered to other electrons, perhaps lifting them above the
ionization threshold or bringing them close enough so that
tunnelling is again possible. If not enough energy is provided
at this stage, perhaps another rescattering process can follow
until eventually multiple ionization takes place or the pulse
disappears. However, before the escape to multiple ionization
all excited electrons pass close to the nucleus where they interact
strongly with the each other and with the Coulomb attraction.
During this phase their (classical) motion is very fast compared to
the field oscillations and an adiabatic analysis, keeping the 
field fixed, can be applied. Morevoer, because of the strong interaction
all memory of the previous motion is lost, so that the initial
state for the multiple ionization event is a statistical distribution
of electrons close to the nucleus. Our discussion starts once this
intermediate cloud of excited electrons has formed. We do not consider
the process by which it has been generated; for instance, 
one might imagine exposing ions to both an electric field and
an electron beam.

The arguments that follow focus on two electron escape but can easily
be extended to discuss multiple ionization, as indicated below.
The Hamiltonian then has as usual the kinetic energy of the electrons, 
their mutual repulsion, the attraction to the core and
the potential due to the oscillating electric field. 
In many experiments the recoil momentum of the ion is measured,
and given the extremely small momenta of the photons it is possible 
to calculate 
it as the sum of the momenta of the electrons.

Initially, there is no field and the atom is in its ground state.
In the final state, after the pulse is turned off, both electrons
are free and have positive total energy. Not all the energy
difference has to be provided by the impacting electron since there
can be additional acceleration by the field after the electrons 
escape from the core region. However, within the adiabatic assumption motivated 
above, the energy content of the intermediate electron cloud has to be 
high enough to let both electrons escape from the immediate vicinity of the 
nucleus. Without field this implies positive energy, but if the field is
on and non-zero, a Stark saddle forms some distance away from the 
nucleus and the electrons can escape over it. The rapid acceleration
downfield will then pull the electrons out and feed in the energy
needed to remain asympotically free once the pulse is gone.

The Stark saddle that forms in the field provides a focus and a 
bottleneck for the electrons which they have to cross in order to leave
the atom. All electrons see the same saddle and would like to cross it,
but if they try to do simultaneously, as suggested by the observed
electron correlations, their mutual repulsion gets in their way.
Suppose that one electron is slightly ahead of the other when running
up the hill towards the Stark saddle: the one that is ahead has the
advantage that the repulsion with the companion pushes it uphill,
whereas the one behind not only has to fight the attraction to the 
nucleus but also the repulsion from the one ahead. Their interaction
is perfectly balanced if they cross the saddle side by side,
with reflection symmetry with respect to the field axis. The previous
considerations suggest that deviations from this symmetric configuration
are amplified and cannot lead to simultaneous ionization. 
The arguments used here are similar to the ones advanced by
Wannier in his analysis of double ionization upon electron impact
\cite{wannier,rau}.

Therefore, we propose that near the threshold for double ionization
the dominant path leading to non-sequential double ionization has both 
electrons escape symmetrically with respect to the field axis.
If more than two electrons are ionized simultaneously the natural 
extension is that they form a regular n-gon in a plane perpendicular 
to the field axis.

\section{Symmetric double ionization}
With the field pointing along the $x$-axis and
two electrons confined to the plane $z=0$ their coordinates 
in the symmetric subspace are
$(x,y,0)$ and $(x,-y,0)$ in position and $(p_x, p_y, 0)$ and
$(p_x, -p_y, 0)$ in momenta. 
The classical Hamilton function for this geometry then is
(in atomic units)
\beq
H(p_x, p_y, x, y, t)= p_x^2+p_y^2 + V(x,y,t)
\eeq
with potential energy
\beq
V(x,y,t) = 
- \frac{4}{\sqrt{x^2+y^2}}
+ \frac{1}{2y} + 2F\, x\, f(t)\, \cos(\omega t+\phi)
\label{potential2}
\eeq
and the pulse shape
\beq
f(t)=\sin^2(\pi t/T_d) 
\eeq
where the duration of the pulse is taken to be four field cycles,
$T_d=8\pi /\omega$.
The rescattering of the electrons leads to a highly excited complex
of total energy $\tilde E$ which every now and then is close to the 
symmetric configuration described by the Hamiltonian (1). Any configuration
on this energy shell (for some fixed time $t$) as well as any phase 
$\phi$ of the field is equally likely, and the experimental 
observations are averages over initial conditions and phases.

\begin{figure}
\centering{\epsfig{file=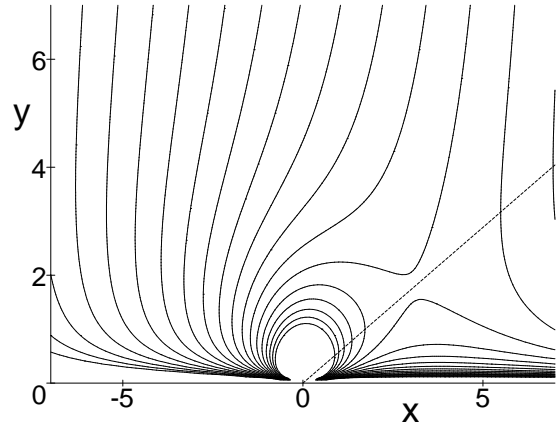%
,scale=0.33,angle=-90
}}
\caption[]{
Adiabatic potential $V(x,y,t)$ for fixed time $t$ 
in the symmetric subspace. The saddle moves along the 
dashed line.
}\label{phspace}
\end{figure}

%
Fig.~\ref{phspace} 
shows equipotential lines for the potential (\ref{potential2}) 
at a maximum
of the field for $F=0.137$~a.u., corresponding to 
an intensity of $6.6\cdot 10^{14}$~W/cm$^2$. The saddle is located
along the line $x=r_s \cos \theta$ and $y=r_s \sin \theta$ with
$\theta = \pi/6$ or $5\pi/6$ and at a distance
$r_s^2=\sqrt{3}/|F\, f(t)\, \cos(\omega t +\phi)|$. For
the above mentioned field the saddle has an energy of 
$V_s=-1.69$~a.u..

\begin{figure}
\centering{\epsfig{file=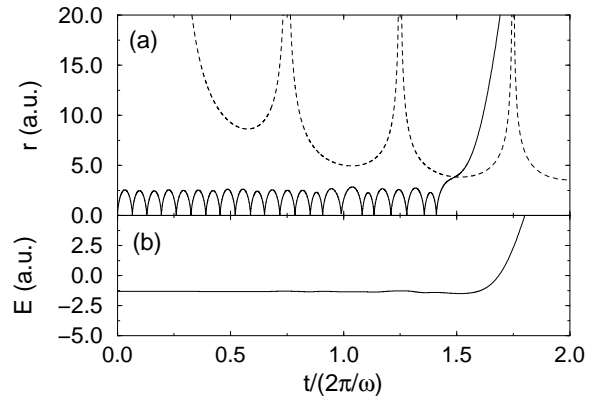%
,scale=0.35,angle=-90%
}}
\caption[]{
Trajectories in the symmetric subspace. One frame shows the
time evolution of the radial distance $r(t)$ (continuous line)
together with the instantaneous
position of the barrier (dashed line) and the other shows the energy.
}\label{traj}
\end{figure}

Sample trajectories within the symmetric configuration are shown in
Fig.~\ref{traj}. It is evident that they cross the saddle during
a maximum of the field and that once on the other side the 
energy increases rapidly. This acceleration is
accompanigned by a rapid separation from the nucleus, so that during
field reversals the electrons will not return to the nuclues and
will remain essentially free.

\begin{figure}
\centering{\epsfig{file=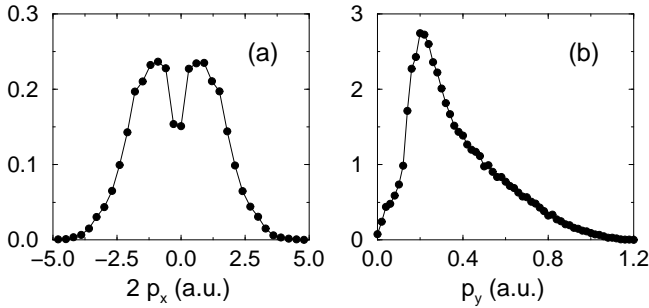%
,scale=0.37,angle=-90%
}}
\caption[]{
Final momentum distributions for $F=0.137$~a.u. and initial energy 
$\tilde E=-0.58$~a.u.:
(a) ion momentum parallel to the field axis and
(b) perpendicular momentum of one electron.
}\label{distrib}
\end{figure}

A quantity of direct experimental interest is the distribution of
momenta of the ion, estimated as ${\bf p}_{\rm Ion} \approx - (
{\bf p}_1 + {\bf p}_2)$
\cite{weber1,weber2,rottke,weber3}. 
The many realization of the experiment
and the unknown details of the initial conditions can be modelled
by averaging over all initial conditions of 
prescribed energy and all phases of the field.
The results for peak field strength $F=0.137$~a.u. are shown 
in Fig.~\ref{distrib}. 
For initial energy $\tilde E=-0.58$~a.u.
the final distribution of momenta clearly shows the double hump
structure also seen in experiment. 
The perpendicular momentum of a single electron 
show a clear suppression near zero
momentum and a long tail. The suppression for small momenta is a clear
sign of electron repulsion. We are not aware of expeperimental
data on this distribution.

\section{The saddle and three-dimensional motions}
Within the symmetric subspace mentioned before the position of the 
saddle separating trapped motion from ionized motion is clear.
And as in many models of chemical reactions it has one unstable
direction that defines the reaction coordinate and a stable motion 
perpendicular to the reaction coordinate. However, in the space
of six degrees of freedom of the full 3-d two electron motion 
and in the adiabatic approximation for the field the
stability analysis of the saddle reveals two unstable directions and
four stable ones. The stable directions are of minor importance:
if excited they persist in the neighborhood of the saddle as some
uncoupled finite amplitude motions. The second unstable direction
besides the reaction coordinate is responsible for the importance of 
the symmetric subspace. Motion leading away from this symmetric
subspace will typically have one electron escaping and the other 
returning to the nucleus. This corresponds to single ionization.
The electron returned to the nucleus may have enough energy
to ionize in the next step or may gain additional energy from the 
field to ionize later. Either way, the electrons do not escape
symmetrically and simultaneously, so that there are no correlations
between the two outgoing electrons and the ionization is sequential.

\begin{figure}
\centering{\epsfig{file=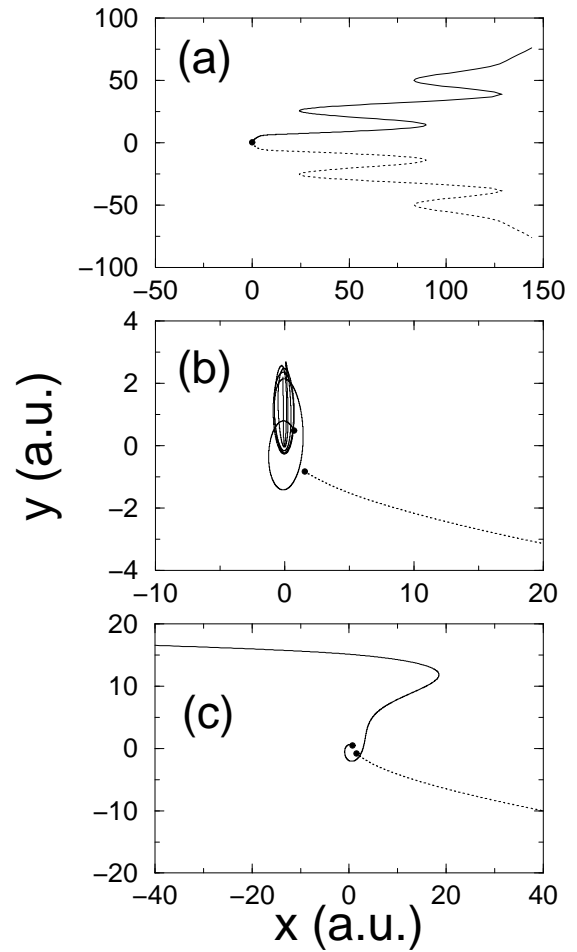%
,scale=0.7,angle=-90%
}}
\caption[]{
Trajectories of electrons outside the symmetric subspace
for $\tilde E=-0.58$~a.u.. Initial 
positions are close to the saddle and marked by heavy dots; the electrons
are distinguished by dotted and continuous tracks. 
Frame (a) shows for reference the ionization in the symmetric subspace.
Frame (b) shows a case where outside this symmetric subspace one
electron escapes and the other falls back to the ion. Frame (c)
shows an example of sequential ionization of both electrons in
opposite directions.
}\label{3dtraj}
\end{figure}

Without going into the technical details of this analysis, we can
illustrate some of these features with trajectories started slightly
outside the symmetry plane (Fig.~\ref{3dtraj}). 
Fig~\ref{3dtraj}a shows 
initial conditions on the saddle and symmetrically escaping
electrons. For some deviation from symmetry, one electron escapes,
the other remains trapped to the nucleus (Fig.~\ref{3dtraj}b). 
It is possible, however, that
the second electron picks up enough energy to ionize itself 
(Fig.~\ref{3dtraj}c). In this figure the loss of correlation between 
the electrons is evidenced by their escape in opposite directions.

\section{Triple and higher ionization}
The model is easily extended to the case of simultaneous
removal of more than two electrons \cite{rottke}. The key assumption
is that the process is dominated by a symmetric configuration of the 
electrons with respect to the field polarization axis. Specifically,
we assume that all electrons move in a plane perpendicular to the
field and that they obey a $C_{Nv}$ symmetry, which generalizes the
$C_{2v}$ symmetry of the previous case.
The reflection symmetry limits the momenta to be parallel to the 
symmetry planes and thus confines the motion to a dynamically allowed 
subspace in the high-dimensional $N$-body phase space. 

With the electric field directed along the $z$-axis the
positions of the $N$-electrons are
$z_i=z$, $\rho_i=\rho$ and $\varphi_i=2\pi i/N$,
where $(\rho_i,\varphi_i,z_i)$ are cylindrical coordinates. 
The momenta of the electrons are all identical, 
$p_{\rho,i}=p_\rho$, $p_{z,i}=p_z$ and $p_{\varphi,i}=0$.
For this geometry the classical 
Hamiltonian for $N$ electrons,
for zero total angular momentum along the field axis
reads
\beq
H(p_\rho,p_z,\rho,z,t)=N
\frac{p_\rho^2+p_z^2}{2}+V(\rho,z,t),
\label{h}
\eeq
with potential energy
\beq
V
=-\frac{N^2}{\sqrt{\rho^2+z^2}}+\frac{N(N-1)}{4\rho\sin (\pi/N)}
+NzFf(t)\cos (\omega t+\phi)\,.
\label{p}
\eeq
The equipotential curves look very much like the ones shown for
two particles and the dynamics is similar. One interesting aspect
of this many electron configuration is that it is limited to at
most 13 electrons: for larger numbers of electrons the repulsion
between overweights the attraction to the nucleus and no
saddle configuration can be found.

\section{Final remarks}
The present considerations suggest that correlated, non-seuqential
multiple ionization in strong laser pulses proceeds through
a saddle configuration with symmetrically moving electrons. The
configurations can be seen analogous to the symmetrically escaping
electrons in double ionization without field as discussed many years
ago by Wannier. As in that case it is possible to derive a
threshold law, which, however, is not only different from his but
also much more difficult to observe because of the presence
of the laser pulse. Further consequences of the model are under
investigation.

Financial support by the Alexander von Humboldt
Foundation and KBN under project 2P302B00915 are gratefully acknowledged.



\narrowtext

\end{multicols}

\begin{references}
\bibitem{koch1995}
P. M. Koch and K. A. H. van Leeuwen, Phys. Rep. {\bf 255}, 289 (1995).

\bibitem{delande1991}
D. Delande, in: {\it Les Houches Session LII, Chaos and Quantum Physics
1989}, editors M. J. Giannoni, A. Voros and J. Zinn-Justin,
(North--Holland, Amsterdam), 665 (1991).

\bibitem{dz1997}
D. Delande and J. Zakrzewski, in: {\it Classical, semiclassical and
quantum dynamics in atoms} H. Friedrich and B. Eckhardt eds., Lecture
notes in physics no.485 (Springer, Berlin) 1997.  

\bibitem{Rost} J.M. Rost, Phys. Rev. Lett. {\bf 72}, 1998 (1994);
Phys. Rep. {\bf 297}, 271 (1999)
  
\bibitem{Kulander1} D.N. Fittinghof, P.R. Bolton, B. Chang,
and K.C. Kulander, Phys. Rev. Lett. {\bf 69}, 2642 (1992)

\bibitem{Kulander2} B. Walker, B. Sheehy, L.F. DiMauro, P. Agostini,
K.J. Schafer, and K.C. Kulander, Phys. Rev. Lett. {\bf 73}, 1227 (1994)

\bibitem{weber1} Th. Weber, M. Weckenbrock, A. Staudte, L. Spielberger, 
O. Jagutzki, V. Mergel, F. Afaneh, G. Urbasch, M. Vollmer, H. 
Giessen and R. D\"orner, Phys. Rev. Lett. {\bf 84}, 443 (2000)

\bibitem{weber2} Th. Weber, M. Weckenbrock, A. Staudte, L. Spielberger, 
O. Jagutzki, V. Mergel, F. Afaneh, G. Urbasch, M. Vollmer, H. 
Giessen and R. D\"orner, J. Phys. B: At. Mol. Opt. Phys. 
{\bf 33}, L1 (2000)

\bibitem{rottke}
R. Moshammer, B. Feuerstein, W. Schmitt, A. Dorn, C.D. Sch\"oter,
J. Ullrich, H. Rottke, C. Trump, M. Wittmann, G. Korn, K. Hoffmann
and W. Sandner, Phys. Rev. Lett. {\bf 84}, 447 (2000)

\bibitem{weber3} Th. Weber, H. Giessen, M. Weckenbrock, G. Urbasch,
A. Staudte, L. Spielberger, 
O. Jagutzki, V. Mergel, M. Vollmer, and R. D\"orner, 
Nature {\bf 405}, 658 (2000).

\bibitem{Corkum} P.B. Corkum, Phys. Rev. Lett. {\bf 71}, 1994 (1993)

\bibitem{Kulander3} K.C. Kulander, J. Cooper, and K.J. Schafer,
	Phys. Rev. A {\bf 51}, 561 (1995)

\bibitem{Kulander4} B. Walker, B. Sheehy, K.C. Kulander, 
and L.F. DiMauro, Phys. Rev. Lett. {\bf 77}, 5031 (1996)

\bibitem{becker1} A. Becker and F.H.M. Faisal, 
J. Phys. B {\bf 29}, L197 (1996)

\bibitem{becker2} A. Becker and F.H.M. Faisal, 
J. Phys. B {\bf 32}, L335 (1999)

\bibitem{Kulander5} B. Sheehy, R. Lafon, M. Widmer, B. Walker,
L.F. DiMauro, P.A. Agostini,
and K.C. Kulander, Phys. Rev. A {\bf 58}, 3942 (1998)

\bibitem{becker3} A. Becker and F.H.M. Faisal, 
Phys. Rev. A {\bf 59}, R1742 (1999)

\bibitem{becker4} A. Becker and F.H.M. Faisal, 
Phys. Rev. Lett. {\bf 84}, 3546 (2000).

\bibitem{wannier}  
{G.H. Wannier}, {Phys. Rev.} {\bf 90}, {817} (1953)

\bibitem{rau}  
{A. R. P. Rau }, {Phys. Rep.} {\bf 110}, {369}, (1984)


    Z. Phys. Chemie B \bf 19\rm, 203 (1932);
    Trans. Faraday Soc. \bf 34\rm 29, (1938)


\end{references}
\end{document}